\documentstyle[10pt,aaspp4]{article}
\newcommand{\ltsim}{\raise 2pt \hbox {$<$} \kern-1.1em \lower 4pt \hbox {$\sim$}}
\newcommand{\gtsim}{\raise 2pt \hbox {$>$} \kern-1.1em \lower 4pt \hbox {$\sim$}}

\lefthead{Giovannini, Taylor, Arbizzani et al.}
\righthead{A Giant Radio Galaxy with Superluminal Motion}

\begin{document}

\title{B2 1144+35: A Giant Low Power Radio Galaxy \\
       with Superluminal Motion}

\author{G. Giovannini\altaffilmark{1,2}, G.B. Taylor\altaffilmark{3}, 
E. Arbizzani\altaffilmark{2}, M. Bondi\altaffilmark{2}, \\ 
W.D. Cotton\altaffilmark{4}, L. Feretti\altaffilmark{2}, 
L. Lara\altaffilmark{5} and T.Venturi\altaffilmark{2}}

\altaffiltext{1}{Dipartimento di Fisica, Universita' di Bologna,via B.Pichat 
6/2, 40127 Bologna, Italy}
\altaffiltext{2}{Istituto di Radioastronomia del CNR, via Gobetti 101, 40129 
Bologna, Italy}
\altaffiltext{3}{National Radio Astronomy Observatory, P.O. Box 0, Socorro NM 
87801, USA}
\altaffiltext{4}{National Radio Astronomy Observatory, 520 Edgemont Rd, \\
Charlottesville VA 22903-2475, USA}
\altaffiltext{5}{Instituto de Astrofisica de Andalucia, CSIC, Apdo 3004, 18080
Granada, Spain}

\centerline{===========================================================}

\centerline{\bf Astrophysical Journal - in press}

\centerline{===========================================================}

\singlespace
\begin{abstract}
We report on centimeter VLA and VLBI observations of the giant, low power 
radio galaxy 1144+35.  These observations are sensitive to structures on 
scales from less than 1 parsec to greater than 1 megaparsec.  Diffuse steep 
spectrum lobes on the megaparsec scale are consistent with an age of $\sim$ 10$^8$ 
years.  On the parsec scale, a complex jet component is seen to move away 
from the center of activity with an apparent velocity 2.7 h$^{-1}_{50}$ c. 
It shows a central 
spine -- shear layer morphology. A faint parsec scale counterjet
is detected and an intrinsic jet velocity of 0.95 c and angle to the line of sight 
of 
25$^\circ$ are derived, consistent with an intrinsically symmetric ejection.  
The central spine in the parsec scale jet is expected to 
move at
a higher velocity and a Lorentz factor $\gamma$ $\sim$ 15 has been estimated
near the core.
The age of this inner VLBI structure is $\sim$ 300 years.  Assuming a constant angle 
to the line-of-sight, the jet velocity is found to decrease from 0.95 c at 20 
mas (32 pc on the plane of the sky) to 0.02 c at 15 arcsec (24 kpc on the 
plane of the sky). 
These findings lend credence to the claim that (1) even the jets of low power 
radio galaxies start out relativistic; and (2) these jets are decelerated 
to subrelativistic velocities by the time they reach kiloparsec scales. 
\end{abstract}

\keywords{radio continuum: galaxies --- galaxies: individual (B1144+35) ---
galaxies: jets --- galaxies: nuclei}

\vfill
\eject

\section{Introduction}

This paper presents new data on a low power radio galaxy 
(B2 1144+35) belonging to a complete sample of radio galaxies which we 
are studying with VLBI data (\cite{g90}).
A major result from this study is the discovery 
that at least in many, and perhaps in all, low power FR I radio galaxies (see 
\cite{fa74}) parsec
scale jets move at relativistic velocities. The derived velocities and 
orientations with respect to the line of sight (see e.g. \cite{lar97} and 
references therein) support
the low-power unified scheme (see e.g. \cite{ur95}) and suggest 
that 
multi-epoch observations should reveal a proper motion in many FR I 
galaxies. Up to now such studies have been carried out only on a small 
number of low power
radio galaxies (e.g. M87, \cite{bir95}; 3C338, \cite{g98a}; NGC\ 315 
\cite{cot98}), 
it is therefore important to examine a larger number of 
FR I radio galaxies with multi-epoch observations to measure jet
proper motions.

The low power radio galaxy B2 1144+35 discussed here, has been identified with 
a faint (m$_{pg}$ = 15.7)
Zwicky galaxy (ZW186.48) in a medium-compact galaxy cluster at a redshift of 
0.0630. An isocontour image taken from the Palomar Observatory Sky Survey 
(POSS) is shown in Fig. 1. The contour image shows that the optical galaxy
has a boxy shape which, according to \cite{bi85}, may
occur in systems that have cannibalized low luminosity galaxies.
A nearby faint companion is embedded in its external region but 
optical spectroscopy is necessary to confirm a genuine connection.
In a recent optical study of bright flat spectrum radio sources, 
\cite{ma96} classify 1144+35 as a BL Lac candidate even though its 
spectrum shows 
H$\alpha$ and [NII] emission lines (\cite{co75,mo92}). From a comparison 
between the measured line
equivalent width and the contrast \cite{ma96} suggest that 
1144+35 could be a diluted
BL Lac. Moreover it was observed in an imaging and spectroscopic survey of
low and intermediate power radio galaxies (\cite{eb89}).
A CCD residual image shows a very
definite arc of dust in the galaxy nuclear region.

The radio galaxy 1144+35 was detected in the X-Ray ROSAT All-Sky Survey 
(\cite{bri95}) with a flux of 7.1 $\times$ 10$^{-13}$ erg cm$^{-2}$
sec $^{-1}$ in the 0.1--2.4 Kev band. \cite{sc99} derived from HRI
ROSAT observations an X-Ray flux density of 8.1 $\times$ 10$^{-13}$ erg 
cm$^{-2}$ sec $^{-1}$ using a powerlaw model and 6.1 $\times$ 10$^{-13}$ erg
cm$^{-2}$ sec $^{-1}$ using a thermal bremsstrahlung model, in the 0.1--2.4 Kev
band.  

  From the radio point of view, 1144+35 has a peculiar structure: \cite{mac98}
classifies it among giant radio galaxies, yet from high 
resolution VLA observations (\cite{fan87}) it has been 
classified a {\it naked} jet source since it shows two faint jet-like 
regions and no extended emission. The radio structure is core dominated 
and the two jet-like features are short and resolved. 
\cite{sc99} observed this source with the Westerbork Synthesis
Radio Telescope (WSRT) at 1.4 GHz and reported WSRT observations at 325 MHz 
from the WENSS survey (\cite{re97}).  They present a detailed image
of the radio core and of two diffuse extended radio structures on either side
of it. The eastern extended lobe shows a leading hotspot and is clearly 
associated with the central source. The northern part of the western radio 
structure is a separate low power radio galaxy; the southern part is a radio 
lobe with an elongated tail most likely associated with 1144+35.
 
The first VLBI observations of this source were carried out by \cite{g90} with
the EVN + Haystack with 2 scans, 13 minutes each. The source was detected, but
the short on source integration time and the poor $u,v$ coverage did not 
allow a discussion of its structure.
It was imaged as part of the Second Caltech-Jodrell survey (\cite{he95}) 
at 5 GHz, where it shows two main knots with some substructure.
A preliminary study of the data from 3 epochs suggested the presence of a 
possible superluminal motion (\cite{g95}).

The new VLA, MERLIN and VLBI data presented in this paper
demonstrate a complex structure over a broad range of
physical scales (1 pc -- 1 Mpc) and confirm superluminal motion.  
We also discuss some
possible explanations for the structure of
this peculiar radio galaxy. We use an Hubble constant H$_0$ =
50 km sec$^{-1}$ Mpc$^{-1}$ which corresponds to a conversion factor of 
1.62 pc/mas.

\placefigure{f1.eps}

\section{Observations and Data Reduction}

\subsection{VLA Data}

On November 1997, we obtained 6 hours of observing time with the VLA of the 
NRAO\footnote{The National Radio Astronomy Observatory is operated by 
Associated
Universities, Inc., under cooperative agreement with the National
Science Foundation} 
in the D configuration at 1.4 and 5 GHz to investigate the large scale 
structure of 1144+35.
The data have been calibrated in the standard way using the NRAO AIPS package 
and imaged using the task IMAGR.
Calibrated data at 1.4 GHz were combined with a short (20 minutes long) VLA 
observation at higher resolution (C array) obtained in September 1997, to have
better angular resolution
while 5 GHz data were combined with a long observation also in the C 
configuration obtained in November 20, 1990 during a VLBI run with the
VLA used as a phased array.  
In this run  
the source was observed for 12 hours, with the array phased every 15 minutes
with a nearby VLA calibrator source.
Since the arcsecond radio core is variable (see Sect. 3.4), a 
fraction of its flux was subtracted from the November 1990 observations
before of combining the two datasets.
The angular resolution and sensitivity of the images presented in this paper
are given in Table 1.

The arcsecond scale radio emission of this source was observed with the VLA
in the A-array configuration at 1.4, 5 and 8.4 GHz, 
on April 1994 and February 1997. Moreover, on September 1997, 1144+35 was 
observed at 15, 22.5 and 43 GHz for the VLA calibrator list.
The source
was observed for about 10 minutes total at each frequency in a few different 
hour angles to obtain better $u,v$ coverage. 
The core flux density was derived by fitting a gaussian to the
nuclear source. In these images due to the lack of short spacings and the short
integration time, only the unresolved arcsecond core is visible and therefore 
we will use these data only to discuss the arcsecond core variability in the 
next sections.

\placetable{arcsecvlamaps}

\subsection{MERLIN Data}

We observed 1144+35 with the MERLIN array on 1995 March 29 at 5 GHz with
a 16 MHz bandwidth for 10 hours. We used the following telescopes: Defford,
Cambridge, Knockin, Wardle, Darnhall, MK2 and Tabley. The data were edited and
amplitude calibrated in Jodrell Bank (JB) using the standard procedure based on
the OLAF programs. 3C286 was used as amplitude calibrator. The data where then
written in FITS format and loaded into AIPS where the phase calibration 
was carried out using
standard MERLIN phase calibrators. The pipeline available in 
Jodrell Bank 
was used as a first step for the phase corrections only. The source was then
mapped in total continuum and polarization and then self-calibrated.

\subsection{VLBI Data}

In Table 2 we summarize the VLBI observations which are discussed in the 
present paper. 
Data marked 1, 2 and 4 are discussed here for the first time, while the
other are {\it literature} data.

The data have been correlated with the Caltech/JPL Block 2 correlator in 
Pasadena (Mk2 VLBI data) and in
Socorro with the NRAO VLBA correlator (VLBA and global VLBI data). Amplitude
calibration was then done using the standard system temperature method in AIPS.
Data were globally fringe fitted and self-calibrated in the standard way.
We made some iterations of phase self-calibration, followed by a final
phase and gain self-calibration when necessary to produce the
final image. Cross-hand correlations were produced for the September 
1997 global
8.3 GHz observations only. Calibration of the instrumental leakage terms was
performed using 4 widely spaced scans on the compact, unpolarized source OQ208
(Roberts et al.\ 1994). The absolute electric vector position
angles were calibrated using two 5 minute scans on 3C279 which has a long-lived
jet component of constant polarization angle (\cite{ta98}). 

\placetable{vlbidata}

Multi-frequency and snap-shot (see note to Table 2) observations were
carried out switching often between frequencies or among different
sources to obtain good and uniform $u,v$ coverage. All other observations
are long integration observations typically of 8 -- 10 hours each.
Mk2 data have been obtained with the old Mark2 system with a 2 MHz
bandwidth and in single polarization mode; VLBA observations are in
single polarization mode with 32 MHz bandwidth, while the global
observation is a dual polarization, 64 MHz bandwidth observation.  In
this latest global observation, we used the following array for the
total intensity images: Effelsberg, Noto, VLBA, and a single VLA antenna. 
Noto was not used for polarization images.

\section{Results}

\subsection{The Megaparsec Scale}

In agreement with \cite{sc99}, we distinguish 3 different
regions in the large scale structure of 1144+35:
the core region, the East, and the West extended emission (see
Figs. 2,3 and 4). The core region coincides with the optical galaxy and
is the dominant feature in all the images.  The core properties will be 
discussed in detail in Sect. 3.4.

\placefigure{f2.eps}

The East extended emission
shows an {\it S} shaped morphology and an uniform low brightness. 
It is clearly connected to the arcsecond core emission even if 
there is a strong discontinuity in the brightness
of the inner nuclear region with respect to the
extended lobe. Two small unresolved regions (see Fig. 4) are
visible at the edge of the extended emission but are
most likely unrelated background sources. The size of this extended
emission is $\sim$6' (580 kpc). 

\placefigure{f3.eps}

The extended emission on the west side can be divided into two
regions: north and south. The northern part is clearly an unrelated
double radio galaxy (see Fig. 4) with a central core and two extended
symmetric lobes. It is identified with a galaxy belonging to
the same group as 1144+35 (\cite{sc99}).
To the south of this radio galaxy we find an extended emission which cannot be
associated with the nearby field galaxy because of the strong
asymmetry and the lack of any direct connection. It is opposite to
the eastern emission of 1144+35 and we
tentatively identify it as the west lobe of 1144+35. 
If 1144+35 is a giant
double radio galaxy, it has an angular size of $\sim$ 13' corresponding
to a projected linear size $\sim$ 1.3 Mpc and a total radio power at 1.4 GHz $\sim$ 1.24 
$\times$ 10$^{25}$ W Hz$^{-1}$.

\placefigure{f4.eps}

To study the spectral index distribution of the extended emission, we
made two images at 1.4 and 5 GHz with $u,v$ coverage as similar as
possible.  To accomplish this we removed the short baselines in the 1.4 GHz 
data and the
long baselines at 5 GHz and obtained two images with a HPBW = 30'' to
enhance the sensitivity of the 5 GHz image to low brightness features.
After correcting for the primary beam attenuation we obtained a
spectral index map between these two frequencies. In Fig. 5 we show
the spectral index map with some contour levels from the 5 GHz image to
show the source structure. The central region of 1144+35 has a flat spectrum
($\alpha \sim$ 0.1 -- 0.2 with $S_\nu$ $\propto$ $\nu^{-\alpha}$). The
east extension is in general very steep; it shows a spectral index
$\alpha \sim$ 0.8 near the core that steepens continuously to the
east, reaching $\alpha \sim$ 2--2.5 in the more external regions. Only
in a small region at RA $\sim$ 11$^h$ 47$^m$ 41$^s$ where a discrete
unrelated source is present we find $\alpha \sim$ 0.9.  
A Comparison of the present result with the spectral index map between 0.3 and 
1.4 GHz discussed by \cite{sc99} shows that the low 
frequency spectrum is flatter than the high frequency one suggesting
the presence of a steepening in the spectrum due to synchrotron losses in an
old radio structure. Future observations, better matched in angular resolution
and $u,v$ coverage, are necessary to confirm this point.
In the SW extended emission we find
$\alpha \sim$ 0.9 where a maximum of the brightness is visible and
\gtsim 1.5 - 2.0 in the extended region. A more detailed analysis is
not possible due to poor sensitivity of the 5 GHz image to the extended
low brightness structure and because of the large primary beam
attenuation.

\placefigure{f5.eps}

We used the 1.4 GHz data and the standard formulae for synchrotron radiation
to calculate the minimum energy density (u$_{\rm min}$) and the equipartition
magnetic field H$_{\rm eq}$.  We assumed a random magnetic field, equally
stored energy in relativistic electrons and heavy particles, a filling factor 
= 1 and a lower and upper frequency cutoff of 10 MHz and 100 GHz respectively.
With these assumptions we found u$_{\rm min}$ $\sim$ 8 $\times$ 10$^{-14}$ erg 
cm$^{-3}$ for the Eastern lobe and u$_{\rm min}$ $\sim$ 7 $\times$ 10$^{-14}$ erg 
cm$^{-3}$ for the Western lobe. The H$_{\rm eq}$ is $\sim$ 1 $\mu$G in both 
lobes. From the derived spectral index of radiating electrons and the magnetic
field present in the Eastern lobe we can estimate the lifetime of radiating 
electrons suffering both synchrotron and inverse Compton losses. According to
the Jaffe-Perola model (\cite{ja74}) that assumes a redistribution 
of electron pitch angles we derive an age in the range 5 to 9 
$\times 10^7$ yrs.

\subsection{The Kiloparsec scale}

We observed 1144+35 with the VLA in C configuration at 5 GHz as part of a VLBI
experiment, obtaining a high quality map at an angular resolution of 3.5''
(see Fig. 6). This image
shows in detail the high brightness central structure visible in lower
resolution images. 
We see a central emission slightly extended in the E-W direction (arcsecond core) 
coincident with the optical
galaxy and faint jets emerging on both sides.  We will call the SE jet
the main jet since it is longer and has a higher surface brightness.
Two faint and extended substructures are visible: J1 and J2, while in
the shorter counterjet only one (CJ1) is visible at about the same
core separation as J1. We measured the j/cj brightness ratio
at 10'' and 15'' from the core and obtained values of 3.6 and 1.1 
respectively. 

\placefigure{f6.eps}

To derive a spectral index map between 1.4 and 5 GHz we used the FIRST
image (\cite{be95}) and an image with same resolution and gridding 
from our 5 GHz data.
Even though the 1.4 GHz data were obtained from a snapshot, we do not expect
missing flux at this frequency because of the small angular size of the 
source.
The main jet has a relatively
flat spectrum: the spectral index is 0.3 and 0.45 in J1 and J2
respectively and 0.2 in between. Also CJ1 has a spectral index $\sim$0.35. 
This result confirms the value derived on larger scales (Fig. 5) 
and the abrupt change in the spectral index
distribution between this arcsecond scale structure and the extended
structure.
The arcsecond core spectral index will be discussed in Sect. 3.4 because of 
the nuclear flux density variability.

\subsection{Polarization Data on Kiloparsec to Megaparsec scales}

The East lobe
shows a high percentage polarized flux (at 1.4 GHz we find 15 to 30\%
in the more external lobe regions, $\sim$ 5\% in the internal lobe
region and up to 60\% near the arcsecond nuclear region). At 5 GHz the
polarization percentage has the same distribution with slightly higher
values (from 30 to 70\%).
The polarization
vectors at 1.4 and 5 GHz are oriented at about the same angle. 

In the extended Western region associated with 1144+35,
polarized flux is only detected in the brighter region. Here the 
polarization vectors are radially oriented and the fractional
polarization is about 15 to 30\% at 1.4 GHz. 
All the W radio emission does not show any change in the
vector's angle between 1.4 and 5 GHz and its depolarization is low.

To better study the central
region we obtained polarization images at 1.4 and 5 GHz with an HPBW of 11'' 
(see Figs. 7,8). At this resolution the signal 
to noise ratio and the angular  
resolution allow us to separate the different components.
At 5 GHz, in the core region, the electric field is oriented at 90$^\circ$ 
but at $\sim$ 15'' from the core it changes 
to 0$^\circ$, and then rotates again to gradually become aligned with
the jet direction at $\sim$ 30'' -- 40'' from the core.
The 1.4 GHz vectors are rotated by $\sim$ 90$^\circ$ in the core,
and the region 15''
from the core is unpolarized.  At 30'' from the core the 1.4 GHz 
electric vectors are aligned with the 5 GHz 
vectors. In the counterjet we see a faint polarized emission oriented at 
$\sim$ 0$^\circ$ at 5 GHz and at $\sim$ 50$^\circ$ at 1.4 GHz.

\placefigure{f7.eps}

\placefigure{f8.eps}

The polarization percentage at 5 GHz is $\sim$ 20\% at 30'' from the core and
decreases slowly to a few percent in the core. At 1.4 GHz we have a comparable
percentage except in the region at 15'' which is completely 
unpolarized at 1.4 GHz.

\subsection{The arcsecond core}

The arcsecond core of 1144+35 is the dominant feature of the radio
emission from this galaxy and has long been known to be variable 
(\cite{ek83}).  Its J2000 position is RA: 11$^h$ 47$^m$ 22.131$^s$
DEC +35$^\circ$ 01' 07.52'' (see 
\cite{jo95} for a more accurate position).
In Table 3 we report the arcsecond core
flux densities available in the literature and the new results obtained by
us, in order to study the flux variability and spectrum. Only core
flux densities from high resolution data have been used so as to 
avoid contamination by the extended emission.  The arcsecond
core flux density is well sampled at 1.4 GHz from 1982 to 1998 and at
5 GHz from 1974 to 1998 while only sparse data at 8.4 and 15 GHz are
available from 1990 onwards. We obtained one observation at very high
frequencies (22 and 43 GHz) to study the high frequency spectrum (see
Figures 9 and 14).

At 1.4 and 5 GHz we see an increase in the 
core flux density from 1974 to 1992 (1.4 GHz), and to 1994 (5 GHz). 
In this time range the core flux density
increased by about a factor of 2. 
After 1980 the flux density variations are smooth on time scales less than a
few years.
Starting in $\sim$ 1990-1992 the core flux density decreases and we observe a 
delay in the decrease at lower frequencies possibly due to opacity effects.

We can derive the core spectral index by
comparing observations at the same time or very near in time.
Using 1997 data we have $\alpha \sim$ 0.43 between 5 and 15 GHz and 0.30
between 15 and 43 GHz with a possible flattening between 1.4 and 5 GHz 
($\alpha \sim$0.13, see Fig. 14). Data in the range 1.4 to 8.3 GHz are 
at the same epoch 
while data in the range 15 to 43 GHz were obtained 6 months apart.
However, the slow variability and the suggestion of a constant spectral
index in time indicates that this short delay should not influence the
radio spectrum. 
The present results, obtained with almost simultaneous observations rule out 
the classification of the arcsecond core as a Gigahertz Peaked Spectrum (GPS)
source suggested by \cite{sn95} and \cite{sc99}. Their conclusion
was based on observations taken at too different epochs given the core flux 
density variability.

\placetable{arcsecflux}

\placefigure{f9.eps}

\subsection{The parsec scale}

\subsubsection{Total intensity and spectral indices}

In Figure 10 we present the full resolution image obtained with the MERLIN array 
at 5 GHz. It shows a 
central source with an extended emission ($\sim$ 100 mas) in the direction
of the main jet and a marginally visible extension also in the opposite
direction.

\placefigure{f10.eps}

With the VLBA at 1.6 GHz the central source is resolved in a double structure
(Fig.~11) with a weak emission
at about 60 mas from the stronger component, in agreement with the
MERLIN image at 5 GHz. 

At 5 GHz with the VLBA, the parsec scale structure is resolved in four main 
substructures:
two compact components (A and B in Fig. 12) separated by 3--4 mas, a discrete
component (C) with a short symmetric extension located at about 20 mas 
from component A and a faint extended emission surrounding A and B components
and clearly elongated in the direction of (C).
The shape of this extended feature is not the same 
in the different epoch and frequency images due to 
different $u,v$ coverage and
sensitivity to low brightness structure, however the inner peak of this 
feature on the top (North) of A and B is always visible and we have named it A1 
in Fig. 12.

\placefigure{f11.eps}

\placefigure{f12.eps}

\placefigure{f13.eps}

In Figure 13 we show our most recent and highest resolution image at 8.4
GHz.  The radio structure is very similar to the 5 GHz image, although
C is clearly resolved into three
components, labeled C, D and E.  Component A1 is resolved
into a complex structure roughly aligned with components C, D and E 
and the kpc-scale jet.  A fairly compact component B1 has emerged
just west of component B.  There is also a clear bridge between
components A and B.

The core variability found at arcsecond resolution implies that 
a flux density variability must be associated with some or all components
detected in the parsec scale.
We have compared the flux density taken at different epochs of components 
A, B, and C
at the same frequency: component C is almost constant at 5 and 1.4 GHz
while at 8.4 GHz it may have increased its flux density (58 mJy on
March 1995; 65 mJy on September 1997).
Components A and B follow the trend of the arcsecond core:
their flux density slightly increases or is constant from 1990-94 and 
decreases in more recent observations. Since the flux density of component A 
($\sim$ 185 mJy
in the last 8.4 GHz map) is 
much higher than the B (30 mJy) and C (65 mJy) flux density it is clear that 
most of the arcsecond core flux density variability can be attributed
to variability of component A.

To derive the spectral index of the different components we used the
VLBA observation obtained on Nov. 26th, 1995 where the source was
observed simultaneously at 1.4, 5, and 8.4 GHz. Moreover we used
the flux densities at 15 and 22 GHz (\cite{he96}) 
which were not too far in time
given that the flux density variability is a smooth function of time
(see Section 3.4). The spectrum of components A, B, and C is shown in Figure 14
where we give also the arcsecond core spectrum at epoch 1997 for a comparison.
We do not show the spectrum of A1 since the flux density of this 
extended feature is strongly influenced by the different $u,v$ coverages of the
observations.

Components A and B show similar spectra: we derive for component
A $\alpha^{1.4}_{8.4}$ = 0.45 and $\alpha^{8.4}_{22.0}$ = 0.92 while component
B has $\alpha^{5.0}_{8.4}$ = 0.38 and $\alpha^{8.4}_{22.0}$ = 0.98.
Component C has an inverted spectrum peaked around 8.4 GHz; 
$\alpha^{1.4}_{8.4}$ = $-$0.34 and $\alpha^{8.4}_{22.0}$ = 0.45 (see Fig. 14).

Based on its spectral properties we identify component C as the center
of activity and the other components as parts of a parsec scale jet.
The core is therefore
self-absorbed with the maximum flux density at $\sim$ 8.4 GHz which
implies a magnetic field of 3--5 Gauss.  The two extended
symmetric regions (D,E) visible at high resolution on both
sides of the core component C are the base of a
two sided symmetric jet structure.  No further jet emission is visible on
the NW side (counter jet side), while on the SE side (the jet side) we
have the extended structure A, B, A1 and B1. Such a jet asymmetry will 
be discussed in Sect. 4.

We identify the extended complex structure A,A1,B,B1 as a clear evidence of a limb 
brightened jet: the maximum surface brightness is on the sides of a 
resolved jet. The projected jet opening angle is of $\sim$ 10$^\circ$. 
Such a structure is expected by the {\it central spine - shear layer} 
jet model (\cite{la96}). 
The image in Fig. 13 shows a gap of emission beyond D  
implying that the shear layer is not visible at the beginning but 
appears at $\sim$ 15 mas from the core.
In this context we
interpret the A, B and B1 structures as due to interaction of the 
parsec scale jets with an inhomogeneous surrounding medium (clouds?). 

The alternative possibility that A1, D and E components are the parsec scale jet, 
as suggested by their alignment with the core, while
A, B and B1 are shocked components ejected at
a difference position angle with respect to the jet direction, appears more 
problematic. The good alignment of components A1, D and E with the core shows that 
this should be
the present jet position angle and therefore A, B and B1 should be an older
emission. However we note that A1 component looks more extended than A and
B components, moreover the estimated velocity of A1 (see Sect. 3.5.3), even if 
affected by
a large uncertainty, is very similar to the A and B measured velocity.
Therefore we interpret the complex parsec scale morphology as a limb
brightened jet with a relevant asymmetry on one side probably due to
a strong interaction with a non symmetric external medium.

\placefigure{f14.eps}

\subsubsection{Polarization on the parsec scale}
 
In Fig. 15 we present the total intensity VLBI image obtained from the Sept. 97 
epoch observation
at 8.4 GHz with polarized intensity vectors.  Components A and B are polarized at 
the $\sim$ 0.4\% and
1.2\% level respectively, while the extended northern structure (A1)
is polarized at a level of $\sim$ 5\%. A fractional polarization of 0.3\%
(170 $\pm$ 30 $\mu$Jy) was found at the core (C) position.  The
polarization angles are not aligned along any preferred
orientation. This could indicate a disorder in the magnetic field
and/or a high and changing Rotation Measure, either one of which is consistent
with the low fractional polarization observed.
This result is in agreement with the presence of a dense interstellar medium
in the parsec scale region.

\placefigure{f15.eps}

\subsubsection{Proper Motion}

We used the observations made at different epochs to measure the apparent
proper motion of components A, A1, B, D, and E with respect to the core
component C.  
First of all we
compared observations obtained at the same frequency at different
epochs to avoid any possible spurious result due to different spectral
indices of components and afterwards we compared the results obtained
using all the data. As expected from the uniform spectral
index and the small size of components we found that the proper motion
is only marginally related to the observing frequency and therefore we
used all the data, except the 1.4 GHz ones where the angular resolution
is too low to properly separate different components. 
In Table 4 and Fig. 16 we show the distance of components A, B, and E from C at 
each epoch. From these data we derive 
that no proper motion is measurable within the errors for  
component E ($\beta$ ~ \ltsim ~ 0.2). Components A and B show a clear
motion from the core C in the direction of the kiloparsec 
scale structure with a constant velocity. The average velocity is 2.78 
$\pm$ 0.1 c for component A and 2.62 $\pm$ 0.1 c for component B. 
The motion of component A1 is not well determined
because of the extended size and low surface brightness of this component. 
Comparing different
epoch data we note that the radio emission centroid of A1 is always in between
A and B components suggesting a similar velocity of A1
with respect to A and B.
For component D no reliable measures were
possible because of the lack of any visible substructure.

\placetable{masdistance}

\placefigure{f16.eps}

\section{Jet Orientation and Velocity}

  From the measured apparent proper motion of the parsec scale jet, 
$\beta_a$ $\sim$ 2.7 h$_{50}^{-1}$,
we can constrain the jet orientation ($\theta$) and velocity 
($\beta$c):

\centerline {$\beta$ = $\beta_a$/($\beta_a cos\theta + sin\theta$)}
 
We find that the 1144+35 jet has to move at a minimum intrinsic velocity
of $\sim$ 0.94 c and that $\theta$ has to be smaller than 40$^\circ$.
The angle $\theta$ corresponding to the minimum velocity is 
$\sim$ 20$^\circ$.

Assuming an intrinsic symmetry in the parsec scale structure of 1144+35, 
we identify the component E as the counterpart of the brightest region of
the main jet, on the counter jet side.
Due to the expected small viewing angle of the source with respect 
to the line
of sight (see above), the size of the counterjet emission is expected to 
appear smaller than that of the jet emission.
Therefore assuming
intrinsic symmetry we can use the observational 
parameters to derive constraints on the velocity and orientation of this
parsec scale jet independently of the Hubble constant as following:

1) Comparing the apparent velocity of the jet $\beta_{\rm aj}$, and of the 
counterjet $\beta_{\rm acj}$, we have (see e.g. \cite{mi94}):

\centerline {$(\beta_{\rm aj} - \beta_{\rm acj}$)/$(\beta_{\rm aj} + \beta_{\rm acj})$ = $\beta cos\theta$}

\vskip 0.5truecm
Given our limit on $\beta_{\rm acj}$~~ \ltsim  ~~0.2, we find: 

\centerline{$\beta cos\theta$ \gtsim 0.86.}

2) From the jet -- counterjet arm ratio (\cite{g98a,ta97}), we derive: 

\centerline{$\beta cos\theta$ $\sim$ 0.82.}

Comparing different constraints, we conclude that the 1144+35 jet has an 
intrinsic high velocity
$\sim$ 0.95c or higher at a small angle with respect to the
line of sight (25$^\circ$ or smaller). 
Given the large size of the extended emission and assuming no large difference
in the orientation from the parsec to the Megaparsec scale structure, 
we will assume a
value of $\theta$ $\sim$ 25$^\circ$. With this orientation with respect to
the line of sight and with $\beta \sim$ 0.95, the Doppler factor $\delta$ is 
$\sim$ 2.25, the intrinsic jet opening angle is $\sim$ 4.2$^\circ$ and
the total linear size of the source becomes $\sim$ 3 Mpc putting it among the
largest of giant radio galaxies.

An independent constraint on $\delta$ can be derived from the 
synchrotron-self-Compton (SSC) model of X-ray emission from the nuclear 
region (see \cite{mar87,ghi93}). In principle when the 
core angular size and the
non-thermal nuclear X-ray emission is known, the comparison 
between the predicted and observed X-ray flux density gives constraints 
on the Doppler factor.
  From the upper limit to the core size from our VLBI data (~\ltsim
flux density measured by \cite{bri95} is
non-thermal nuclear emission, we 
derive a lower limit for $\delta$ (see also \cite{g94}). Taking 
8.4 GHz as the core self-absorption frequency we find a Doppler factor
$\delta$ larger than 
1.4 in agreement with our previous estimate.

  From the estimated intrinsic velocity of the parsec scale jet, we can derive 
that the
A+B+A1+B1 complex was ejected from the central engine about 300 years ago in
the source reference frame.  
In this context the D feature is a more recent ejection from the core whose
proper motion is not yet visible due to the lack of
visible sub-structures.  We would predict that in the near future it should 
be possible to see this component moving away from the core.
The velocity obtained above refers to the brightest jet region (the shear
layer). The presence of a central jet region (spine) with a lower brightness
can be due to a higher velocity of the inner jet region with respect to
the brighter external jet regions. A higher velocity implies in our case 
a low Doppler factor and therefore the jet central spine
moving at high velocity will be less visible than the external shear layer.
Also
the presence of a gap of emission between D and the extended jet emission
could be due to a low Doppler factor of a fast moving jet.
For the A and B structures (but A1 and B1 are very likely moving at the same 
velocity) we have derived a doppler factor $\delta$ $\sim$ 2.25. 
This value implies that the intrinsic radio emission is amplified of a 
factor 7.5 -- 8. To justify the dimming
in the surface brightness in the central region of the jet and the gap, we 
have to assume that the fast jet spine should move with a Lorentz factor 
$\gamma$ $\sim$ 15.
A value of $\gamma$ $\sim$ 15 corresponds to a velocity $\sim$ 0.998 c and a 
Doppler factor $\delta$ $\sim$ 0.698 which implies that the intrinsic radio
emission is dimmed by a factor $\sim$ 0.4.
This model assumes an intrinsically symmetric jet
while our images show a clear difference between components A, B, B1 and A1. As
discussed in Sect. 3.5.1 we assume that these inhomogeneities are due to the
interaction of the external jet regions with an inhomogeneous and
clumpy surrounding medium. If this is the case, the inner jet spine should
not be affected by the interaction with the external medium.
  
A free expanding jet is expected to show an intrinsic opening angle 
$\sim 1/\gamma$, (see e.g. \cite{sa98}) while a well confined and 
collimated jet will show a smaller
opening angle. For the present source the estimated value of $\gamma$ $\sim$ 15
(1/$\gamma$ $\sim$ 4$^\circ$) is in very good agreement with the intrinsic
jet opening angle derived before (4.2$^\circ$ if $\theta \sim 25^\circ$).

In this scenario, at the jet beginning there is no or very little
difference in velocity between the spine and the shear layer therefore the 
whole jet is doppler deboosted
and we have a gap in the radio emission. At about 15 mas from the 
core (corresponding
to a de-projected distance of $\sim$ 55 pc) the external jet regions are moving
at a lower velocity because of their interaction with the surrounding medium
and the shear layer becomes more visible
being Doppler boosted, while the inner jet region 
is still moving at high velocity and consequently is de-boosted. 
The presence of component D near the core could have two possible
explanations: (1) the jet is accelerating as found in NGC 315
(\cite{cot98}); or 
(2) the intrinsic brightness of the jet near the core is high
and therefore component D is visible despite the dimming due to the 
de-boosting effect.

With the existing data we cannot exclude that
a transversal asymmetry causes: (1) the difference between the A, B, B1 and
A1 regions; (2) the low brightness in the jet central region; and (3) the gap
near the core. Future multi-frequency observations of the polarized emission
in the jet should help to clarify the degree of
interaction between the jet and the surrounding medium.  Furthermore, long
term monitoring of the source should allow the detection of a proper motion in
the counterjet side (component E). This measurement is important since it
will constrain the intrinsic jet velocity and allow for a direct
distance determination.

We found a similar parsec scale morphology with
space VLBI observations of the BL-Lac
object Mkn 501 (\cite{g98b}): in this source we see a 
centrally peaked jet moving
at a velocity in the range 0.990 -- 0.999c in the inner 40 -- 45 pc 
from the core, which becomes limb brightened at a de-projected distance 
of $\sim$ 45 pc. 
This result reinforces the scenario of a jet which appears 
limb-brightened at some distance from the core ($\sim$ 40 -- 50 pc) due to the 
brightness of the
shear layer which has slowed down  
with respect to the central spine
because of the interaction with the surrounding medium.

Assuming a constant value for $\theta$, from the j/cj ratio on the kiloparsec
scale we find a jet velocity of $\sim$ 0.28 c at 10'' from the
core and of 0.02 c at 15'' from the core in agreement with the expected jet
velocity decrease in low power radio galaxies (see e.g. 3C449, \cite{fer99}). 
Such a velocity cannot explain
the armlength asymmetry visible at arcsecond resolution (see Fig. 6);
we suggest in agreement with \cite{sc99} that the arcsecond 
arm ratio asymmetry is due to the interaction with an
inhomogeneous external medium which is responsible also for the asymmetric
Megaparsec structure of 1144+35.

\section{Arcsecond Core Flux Density Variability}

The arcsecond core flux density of 1144+35 increases from 
1974 to 
$\sim$ 1994 and decreases from 1994 to 1998. As discussed before, the flux density 
variability is not due to
the activity of the mas core but to changes in the A, B, A1, B1 
complex, specially in the A component being the region with highest 
flux density on the parsec scale.

A model to explain such a flux density turn-over in   
GPS sources was presented by \cite{sn98} and in agreement with 
{\cite{sc99} it could explain the core flux density variability in the 
present source.
In fact, according to the scenario discussed in Sect. 3.5.3 
the flux density variation of a factor
2 between 1974 and 1994 can be related to a slowing down of the shear layer 
of the parsec 
scale jet which increases the Doppler factor and therefore the observed flux.
To produce a factor 2 of increasing flux the Doppler factor had to 
be $\sim$ 1.7 taking into account that now it is estimated to be $\sim$ 2.25. 
Such a variation can be obtained if the velocity decreases from $\sim$ 0.98c 
to 0.95 c in the last $\sim$ 15 mas behind A.

The faster flux density increase from 1974 to 1980 followed by a slower 
increase from
1980 to 1994 (Fig. 9) implies a variable rate of the Doppler factor 
probably due to a velocity decrease not constant in time but with a stronger 
deceleration at the beginning.

The decreasing flux from 1995 to present days could be due to adiabatic 
losses if A 
(and B) are expanding. In an expanding component we expect a delay in the
flux density 
decrease at lower frequencies due to opacity effects, as found in
the 1144+35 core (see Fig. 9). 

\section{The Connection between the Large and Small Scale Structure}

The large scale structure shows a clear discontinuity between the extended
relaxed lobes and the high brightness arcsecond core and jets (Sect. 3.1).
This discontinuity is confirmed by the spectral index map which shows a
large change in the spectral index moving from the arcsecond jet
to the extended East lobe.
These observational data suggest two different phases in the life of
this radio source: the extended structure is a relic emission with an age  
in the range 5 to 9 $\times$ 10$^7$ yrs as estimated in Sect. 3.1, while
the emission on the arcsecond scale has a shorter dynamical age:
assuming an average velocity of \gtsim 0.02 c for the arcsecond 
structure
we derive an age of \ltsim 1.0 $\times 10^7$ yr, taking into account projection 
effects.
We note that the Megaparsec scale East lobe of this source shows similar 
physical properties and is very similar in shape to the extended
relic source 1253+275, found at the periphery of the Coma cluster (Giovannini et
al.\ 1991). We can speculate that if the core of 1144+35 had ceased its
activity, the extended emission could no longer be recognized as due to the 
activity of the galaxy ZW186.48 but it would be considered a relic emission in
a group of galaxies.

A merger event with a gas rich object as suggested also by the {\it boxy}
shape of the optical image could be the origin of the restarted activity.
The large number of {\it young} CSS or CSO sources showing evidence of a recent
merger event (\cite{o'98}) confirms the correlation between these processes
and the radio activity.

Moreover we note that the parsec scale jet emission does not decrease smoothly 
in our VLBI maps. The parsec scale jet appears to stop at $\sim$ 24 mas from 
the core even if we have some evidence from the MERLIN and the 20 cm VLBA map 
that a low level flux density emission
is present at larger distance from the core, in the same position angle. 
This can be due to a dramatic 
expansion of the radio jet after the position of the A component with a 
correspondingly large decrease in the
surface flux density, similar to the jet expansion visible in Mkn 421 and 
Mkn 501 (Giovannini et al.\ 1998b). Alternatively the A, A1, B, B1 complex 
could be a 
strong enhancement in the jet brightness due to a 
strong increase of the nuclear activity.
This burst
should have taken place about 300 yrs ago in
the source reference frame, corresponding to $\sim$ 45 yrs in our reference 
frame. This hypothesis is in agreement with the core flux density variability
discussed in the previous sections.

\section{Conclusions}

We present new VLA and VLBI observations of the low power radio galaxy 
1144+35. These observations allow us to study and discuss the properties of 
this source from the Megaparsec to the
parsec scale. 1144+35 is one of the few sources with observed two-sided 
parsec scale 
jets. These sources are of particular interest since they can be used to 
place constraints on the 
cosmological parameters H$_0$ and q$_0$. Moreover this source is oriented at 
$\sim$ 25$^\circ$ with respect to the line of sight therefore, in agreement 
with unified scheme models, presents properties in between FR I radio galaxies 
and BL-Lac type objects.

On the Megaparsec scale 1144+35 appears to be a giant radio galaxy: 
1.3 Mpc on the 
plane of the sky corresponding to a deprojected linear size $\sim$ 3 Mpc. 
It shows an East lobe with an 
elongated shape, connected
to the core emission and a Western extended emission interpreted as the West 
lobe of a double
radio galaxy. The age of this emission is estimated to be 5 to 9 $\times$ 
10$^7$ yrs.

At kiloparsec resolution, a core and a two-sided jet are present. 
The surface 
brightness and the spectral index distribution suggest a strong discontinuity 
between this structure
and the Megaparsec scale emission. This observational result could be due
to a change in the jet direction or to renewed activity in the galaxy core
probably triggered by a galaxy merger.

The arcsecond core is the dominant feature of this source at arcsecond 
resolution. Its flux density showed a large increase from 1974 to 1980 
(from 280 to 500 mJy at 5 GHz),
followed by a smooth increase until 1992. From 1992 to date the core flux 
density has decreased; the flux density decrement appeared first at higher 
frequencies. The core spectrum obtained from same epoch observations is flat 
and we rule out the
identification of the 1144+35 core as a GPS source.

At parsec resolution we identified the core source (C) and a two-sided jet 
emission very 
asymmetric in shape and properties. This asymmetry, when interpreted in the 
light of unified scheme models and of relativistic jets, constrains the jet
velocity to be
\gtsim 0.95c with an orientation with respect to the line of sight 
\ltsim 25$^\circ$.
These results are in agreement with the detected proper motion with $\beta_a$ 
$\sim$ 2.7.
The source orientation explains its properties intermediate between a FR I 
radio galaxy
and a BL-Lac type object.

The parsec scale jet is limb brightened and its morphology is in agreement 
with the presence of a fast inner jet spine ($\beta \sim$ 0.998) surrounded 
by a shear layer in which the velocity 
decreases because of the interaction with the surrounding medium. 
The estimated age of this structure is of $\sim$ 300 yrs in the source 
reference frame.

\acknowledgements}

We thank R. Fanti, D. Dallacasa and G. Brunetti for critical reading of the 
manuscript and useful discussions.
The Digitized Sky Survey was produced at the Space Telescope Science Institute
under US Government grant NAG W-2166.
This work was partly supported by the Italian Ministry for University
and Research (MURST) under grant Cofin98-02-32

\clearpage

\begin{table}
\begin{center}
\caption{VLA image parameters \label{arcsecvlamaps}}
\medskip
\begin{tabular}{ccclcc}
\tableline  \tableline

Array & Frequency & HPBW & noise & Figure & notes  \\
      &   GHz~~~~~& $\prime\prime$ ~~~~& mJy/b &  N.    &        \\
\hline
C+D   &  1.4      &  20  & 0.06  &  2     & I map   \\
C+D   &  4.9      &  20  & 0.015 &  3     & I map   \\
C+D   &  4.9      &  20  & 0.012 &  3     & P map   \\
C+D   &  4.9      &  11  & 0.014 &  4     & I map   \\
C     &  4.9      &  3.5 & 0.06  &  6     & I map   \\
\tableline
\end{tabular}
\medskip \\
\end{center}
\tablecomments{
See Figure captions for more details; I map is the total intensity image; P map
is the Polarized image used to draw E-vectors superimposed to the total 
intensity images.}
\end{table}

\clearpage

\begin{table}
\begin{center}
\caption{History of VLBI Observations \label{vlbidata}}
\medskip
\begin{tabular}{cccccc}
\tableline  \tableline
Epoch  &  Array &  Array &  Array &  Array &  Array \\
mm-yy  & 1.4GHz & 5.0GHz & 8.3GHz & 15.3GHz& 22.2GHz \\
\hline
11-90  &        & Mk2$^1$&        &        &        \\
11-92  & Mk2$^1$&        &        &        &        \\
06-93  &        & Mk2$^5$&        &        &        \\
01-95  &        &        &        &  VLBA$^3$&        \\
03-95  & VLBA$^2$&        & VLBA$^2$   &        &        \\
08-95  &        & VLBA$^4$   &        &        &        \\
11-95  & VLBA$^1$   & VLBA$^1$   & VLBA$^1$   &        &        \\
02-96  &        &        &        &        & VLBA$^3$   \\
08-96  &        & VLBA$^4$   &        &        &        \\
09-97  &        &        & GLOBAL$^1$ &        &        \\
\tableline
\end{tabular}
\medskip \\
\end{center}
\tablecomments{
1 -- Full synthesis (8-10 hours long) observations; 2 -- 5$\times$6 min. 
snap-shots; 3 -- see D. Henstock, 1996; 4 -- 8$\times$6 min.
snap-shots with 8 MHz bandwidth; 5 -- see Henstock et al. 1995.}
\end{table}

\clearpage

\begin{table}
\begin{center}
\caption{Arcsecond core flux densities \label{arcsecflux}}
\medskip
\begin{tabular}{cccccccc}
\tableline  \tableline

Epoch & Flux(mJy) & Flux(mJy) & Flux(mJy) & Flux(mJy) & Flux(mJy) & Flux(mJy) &Ref. \\
mm-yy & 1.4 GHz & 5.0 GHz & 8.3 GHz & 15 GHz & 22 GHz & 43 GHz &      \\
\hline
01-73 & 340     &   -     &   -     &    -   &   -    &   -    &  1   \\
01-74 &  -      &  290    &   -     &    -   &   -    &   -    &  1   \\
12-74 &  -      &  310    &   -     &    -   &   -    &   -    &  2   \\
02-75 &  -      &  310    &   -     &    -   &   -    &   -    &  2   \\
04-75 &  -      &  330    &   -     &    -   &   -    &   -    &  2   \\
12-75 &  -      &  340    &   -     &    -   &   -    &   -    &  2   \\
08-76 &  -      &  395    &   -     &    -   &   -    &   -    &  2   \\
01-77 &  -      &  425    &   -     &    -   &   -    &   -    &  2   \\
02-80 &  -      &  500    &   -     &    -   &   -    &   -    &  2   \\
08-82 & 529     &   -     &   -     &    -   &   -    &   -    &  3   \\
12-84 & 568     &   -     &   -     &    -   &   -    &   -    &  4   \\
05-85 & 568     &   -     &   -     &    -   &   -    &   -    &  5   \\
02-90 &  -      &   -     &  501    &    -   &   -    &   -    &  6   \\
11-90 &  -      &  551    &   -     &    -   &   -    &   -    &  pp  \\
10-91 & 600     &   -     &   -     &   359  &   -    &   -    &  7   \\
04-94 & 575     &  553    &  450    &    -   &   -    &   -    &  pp  \\
08-94 &  -      &  537    &   -     &    -   &   -    &   -    &  8   \\
07-95 & 570     &   -     &   -     &    -   &   -    &   -    &  8   \\
02-97 & 569     &  483    &  393    &    -   &   -    &   -    &  pp  \\
08-97 &  -      &   -     &   -     &   293  &  262   &  215   &  pp  \\ 
08-97 & 541     &   -     &   -     &    -   &   -    &   -    &  9   \\
\tableline
\end{tabular}
\medskip \\

\end{center}
\tablecomments{
References: 1) Colla et al.\ 1975; 2) Ekers et al.\ 1983; 3) Parma et al.\ 
1986; 4)
Fanti et al.\ 1986; 5) Fanti et al.\ 1987; 6) Patnaik et al.\ 1992; 7) 
Snellen et al.\ 1995; 8) Taylor et al.\ 1996; 9) Schoenmakers et al.\  1999;
pp) present paper.}
\end{table}

\clearpage

\begin{table} 
\begin{center}
\caption{Distances from the core at various epochs \label{masdistance}}
\medskip
\begin{tabular}{cccccc}
\tableline  \tableline
\\
Epoch    & E  &   A   &   B   &  Frequency \\
dd/mm/yy & mas  &  mas  & mas   &    GHz     \\
\\
20/11/90 &  -   &  20.2 &  16.4 &  5.0 \\
10/06/93 & 2.3  &  21.4 &  18.0 &  5.0 \\
27/01/95 &  -   &  22.5 &  18.5 & 15   \\
22/03/95 & 2.1  &  22.5 &  18.6 &  8.4 \\
25/08/95 & 2.2  &  22.5 &  18.6 &  5.0 \\
26/11/95 & 2.4  &  22.7 &  18.8 &  5.0 \\
26/11/95 & 2.2  &  22.8 &  18.9 &  8.4 \\
18/02/96 &  -   &  23.2 &  19.3 & 22  \\
22/08/96 & 2.0  &  23.0 &  19.1 &  5.0 \\
27/09/97 & 2.4  &  23.8 &  19.8 & 8.4 \\

\tableline
\end{tabular}
\medskip \\

\end{center}
\tablecomments{
Positional uncertainties are $\pm$ 0.2 mas for the faint component
E and $\pm$ 0.1 mas for A and B components ($\pm$ 0.07 mas 
at 15 and 22 GHz).} 
\end{table}

\newpage

\figcaption[f1.eps]{Optical image of 1144+35 from the Digitalized Palomar
Sky Survey \label{f1.eps}}

\figcaption[f2.eps]{VLA image at 1.4 GHz. The HPBW is 20'' and the noise 
level is 0.06 mJy/beam. The peak flux is 539.4 mJy/beam and contour levels are:
-0.2 0.15 0.3 0.5 0.7 1 1.5 2 3 5 7 10 30 50 100 300 mJy/beam. 
\label{f2.eps}}

\figcaption[f3.eps]{VLA image at 4.9 GHz. The HPBW is 20'' and the noise 
level is 0.015 mJy/beam. Contour levels are: 0.05 0.1 0.2 0.3 0.5 0.7 1 3
5 10 30 50 100 200 mJy/beam. Lines are proportional to the polarized
intensity and are oriented as the Electric field. \label{f3.eps}}

\figcaption[figure=f4.eps]{VLA image at 4.9 GHz superimposed on an optical
image from the Digitized POSS. The HPBW is 11'' and the noise level 0.014 
mJy/beam. The peak flux is 442.28 mJy/beam and contour levels are: $-$0.08 0.05
0.1 0.15 0.2 0.3 0.4 0.5 0.7 1 2 4 6 8 10 30 50 100 200 300 400 mJy/beam.
\label{f4.eps}}

\figcaption[f5.eps]{Spectral index map between 1.4 and 5 GHz. The grey 
scale is in milli-spectral index. Contour levels are from the 5 GHz image. The
HPBW is 30''. \label{f5.eps}}

\figcaption[f6.eps]{Total intensity image at 5 GHz; the HPBW is 3.5''. 
Contour levels are:
0.2 0.5 1 2 5 10 50 100 300 mJy/beam. The noise level is 0.06 mJy/beam. Jet
substructures are labelled according to Sect. 3.2. \label{f6.eps}}

\figcaption[f7.eps]{Total intensity image at 1.4 GHz with E field polarization vectors
superimposed proportional to the polarized flux density; the HPBW is 11''.
Contour levels are:
0.5 1 3 5 10 30 50 100 300 mJy/beam. \label{f7.eps}}

\figcaption[f8.eps]{Total intensity image at 5 GHz with E field polarization 
vectors
superimposed to the polarized flux density; the HPBW is 11''. Contour levels
are:
0.05 0.1 0.15 0.2 0.3 0.5 1 3 5 10 30 50 100 200 300 mJy/beam. \label{f8.eps}}

\figcaption[f9.eps]{Arcsecond core flux densities at different epochs. 
\label{f9.eps}}

\figcaption[f10.eps]{Full resolution MERLIN image of 1144+35 at 5 GHz. The 
HPBW is 49 $\times$ 42 mas
in PA 50$^\circ$. The noise level is 0.05 mJy/beam. Contour levels are:
$-$0.3 0.3 0.5 0.7 1 3 5 10 30 50 100 200 300 400 mJy/beam. \label{f10.eps}}

\figcaption[f11.eps]{VLBA image at 1.4 GHz of 1144+35. The HPBW
is 11 $\times$ 5 mas in PA 4$^{\circ}$.
The noise level is 0.25 mJy/beam. Contour levels are: $-$0.5 0.5 1
1.5 2 3 5 10 20 30 50 100 150 200 350 mJy/beam. \label{f11.eps}}

\figcaption[f12.eps]{VLBA image at 5 GHz of 1144+35. The HPBW
is 2.9 $\times$ 2.0 mas in PA -2$^{\circ}$.
The noise level is 0.4 mJy/beam. Contour levels are: $-$1 0.6 1 1.5 2 3 5 7 10
20 30 50 100 200 mJy/beam. Different components are labelled according to
Sect. 3.5.1. \label{f12.eps}}

\figcaption[f13.eps]{VLBI image at 8.4 GHz of 1144+35. The HPBW
is 1.45 $\times$ 0.66 mas in PA $-$13$^{\circ}$.
The noise level is 0.05 mJy/beam. Contour levels are: $-$0.2 0.15 0.3 0.5 0.7
1 2 3 5 7 10 20 30 50 100 mJy/beam. Different components are labelled
according to Sect. 3.5.1. \label{f13.eps}}

\figcaption[f14.eps]{Radio spectrum of the arcsecond core (" - open circles) 
from VLA data, 1997 epoch and
of the components A, B and C (VLBI Nov. 26th, 1995 epoch). \label{f14.eps}}

\figcaption[f15.eps]{Same as Fig. 13, with E field polarization vectors 
superimposed proportional to the polarized flux density. Contour levels are:
$-$0.2 0.2 0.5 1 1.5 2 3 5 7 10 30 50 100 mJy/beam. \label{f15.eps}}

\figcaption[f16.eps]{Core distance versus time of the A, B and E components.
Triangles refer to 5GHz data, squares to 8.4 GHz data and hexagons to 15 and 22
GHz data.
Positional uncertainties are not drawn for clarity; they are:
$\pm$ 0.2 mas for component E
and $\pm$ 0.1 mas for A and B components ($\pm$ 0.07 for 15 and 22 GHz data).
Note that the {\it y axis} is not continuous. \label{f16.eps}}

\end{document}